# Design of the Life Signature Detection Polarimeter LSDpol

Christoph U. Keller[*a], Frans Snik[a], C. H. Lucas Patty[a,b], Dora Klindžić[a,c], Mariya Krasteva[a], David S. Doelman[a], Thomas Wijnen[a,c], Vidhya Pallichadath[c], Daphne M. Stam[c], Brice-Olivier Demory[b], Jonas G. Kühn[b], H. Jens Hoeijmakers[b], Antoine Pommerol[b], Olivier Poch[d]

[a]Leiden Observatory, Leiden University, PO Box 9513, 2300 RA Leiden, The Netherlands;
[b]Physikalisches Institut, Universität Bern, Sidlerstrasse 5, 3012 Bern, Switzerland;
[c]Faculty of Aerospace Engineering, Delft University of Technology, Kluyverweg 1, 2629 HS Delft, The Netherlands; [d]Université Grenoble-Alpes, CNRS, IPAG, 38000 Grenoble, France

## ABSTRACT

Many biologically produced chiral molecules such as amino acids and sugars show a preference for left or right handedness (homochirality). Light reflected by biological materials such as algae and leaves therefore exhibits a small amount of circular polarization that strongly depends on wavelength. Our Life Signature Detection polarimeter (LSDpol) is optimized to measure these signatures of life. LSDpol is a compact spectropolarimeter concept with no moving parts that instantaneously measures linear and circular polarization averaged over the field of view with a sensitivity of better than $10^{-4}$. We expect to launch the instrument into orbit after validating its performance on the ground and from aircraft.

LSDpol is based on a spatially varying quarter-wave retarder that is implemented with a patterned liquid-crystal. It is the first optical element to maximize the polarimetric sensitivity. Since this pattern as well as the entrance slit of the spectrograph have to be imaged onto the detector, the slit serves as the aperture, and an internal field stop limits the field of view. The retarder's fast axis angle varies linearly along one spatial dimension. A fixed quarter-wave retarder combined with a polarization grating act as the disperser and the polarizing beam-splitter. Circular and linear polarization are thereby encoded at incompatible modulation frequencies across the spectrum, which minimizes the potential cross-talk from linear into circular polarization.

## 1. INTRODUCTION

Homochirality, the preferential handedness of some biological molecules, provides an exciting opportunity to remotely detect the presence of life. Light reflected off biological surfaces will therefore often be slightly circularly polarized at the $10^{-4}$ level[1]. The reflected light will also be linearly polarized at the $10^{-1}$ level due to surface (Fresnel) reflection. Measuring a small amount of circular polarization on top of a large linear polarization signal is a major challenge because instrumental crosstalk from linear into circular polarization will typically create artificial circular polarization signals at the $10^{-3}$ level.

Over the last few years we have been working on instrumental developments to efficiently and accurately measure circular polarization as a function of wavelength in the presence of large linear polarization signals. The first of these instruments, TreePol[2], used a ferro-electric liquid crystal polarization modulator and a spinning half-wave plate to remove the linear polarization. Both the polarization modulator and the spinning halfwave plate are major complications for a simple, space-qualified instrument. The liquid crystal requires temperature control, and moving parts significantly increase the cost of an instrument and may be the source of unwanted mechanical vibrations. This led us to develop LSDpol, the Life Signature Detection polarimeter, which does not require temperature control and does not have any moving parts. We presented initial ideas in[3]. Sparks et al.[4] independently developed a similar concept. Here we provide a description of the design considerations, the prototype design, initial experimental results and lessons learned for future generations of the LSDpol instrument concept.

---

[*] keller@strw.leidenuniv.nl; home.strw.leidenuniv.nl/~keller/

# 2. POLARIMETRY FIRST, SPECTROSCOPY SECOND

The overarching principle of our design has to be 'Polarimetry First'. In many instrumental designs polarimetry is an afterthought, which limits the sensitivity and accuracy with which polarization can be measured. Since homochirality typically manifests itself in a strong wavelength dependence of the circular polarization signal, the second most important instrumental aspect has to be spectroscopy, hence 'Spectroscopy Second'. In the following we will discuss our approaches to polarimetry and spectroscopy. Angular resolution is not a consideration at this time, although we are investigating the potential for integral-field spectro-polarimetry with an LSDpol-like instrument.

## 2.1 Polarimetry

There are many approaches to spectro-polarimetry. Since a polarization measurement always requires two or more intensity measurements, one has to carefully consider how these intensity measurements are acquired; we will generally refer to *polarization modulation* as the optical element that changes to measure two or more polarization states to combine into a polarization measurement. Since polarimetry has priority, we need to modulate the polarization of light with the very first element. Any other optics such as lenses and mirrors would at a minimum introduce cross-talk from linear to circular polarization if not create linear and potentially circular polarization from unpolarized light. Such an approach has been used successfully to measure the fractional polarization of integrated sunlight at the $10^{-7}$ level[5].

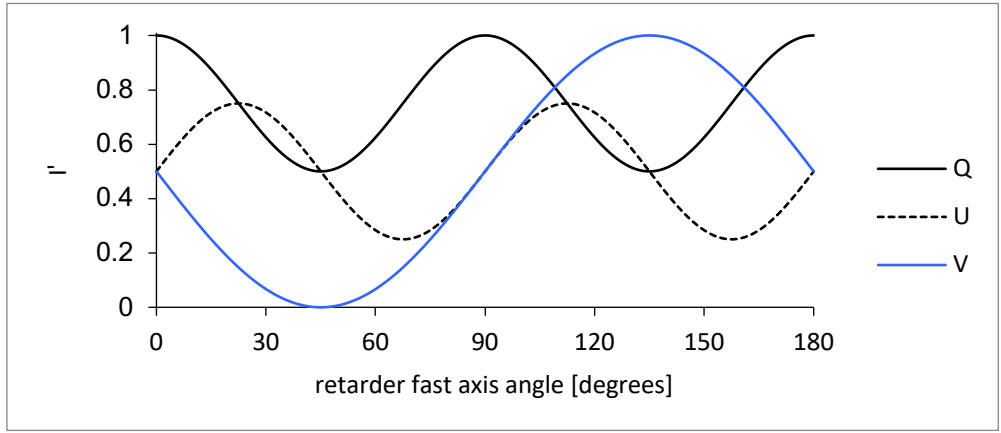

Figure 1. The intensity after a rotating quarter-wave retarder and a linear polarizer when the light source is fully polarized with Q=I, U=I, or V=I.

The second, equally important requirement on our polarization modulation approach concerns the cross-talk from linear to circular polarization. Since we have to expects tens of percent of linear polarization, our instrumental cross-talk has to be at the $10^{-3}$ level or below. As such, it is beneficial to have a modulation approach that inherently has no cross-talk from linear into circular polarization. One approach are piezo-elastic modulators (PEMs) where linear polarization is modulated at twice the temporal frequency of circular polarization[5]. PEMs are relatively fragile and require detectors that operate at tens of kHz, which precludes the use of standard imaging detectors. Another inherently cross-talk free approach is a rotating retarder followed by a linear polarizer; again linear polarization is modulated at twice the frequency of circular polarization. The intensity measured by a perfect rotating retarder polarimeter is given by

$$I'(\theta) = \frac{1}{2}\left(I + \frac{Q}{2}\left((1 + \cos\delta) + (1 - \cos\delta)\cos 4\theta\right) + \frac{U}{2}(1 - \cos\delta)\sin 4\theta - V\sin\delta\sin 4\theta\right),$$

where $I$, $Q$, $U$ and $V$ are the four Stokes parameters of the incoming light, $I'$ is the measured intensity as a function of the fast-axis orientation angle $\theta$, and $\delta$ is the retardation of the retarder. Stokes $Q$ and $U$ describe linear polarization while Stokes $V$ is the difference between left and right circularly polarized light. Values for the polarized components $Q$, $U$ and $V$ are typically expressed as fractions of Stokes $I$. Since we do not want any moving parts, we implement the equivalent modulation with a patterned quarter-wave retarder where the fast axis rotates in one dimension. Such a patterned retarder can be implemented with liquid-crystal technologies[6]. Figure 1 shows the measured intensity as a function of the retarder fast axis orientation in the case of a perfect quarter-wave retarder, which provides 100% modulation for Stokes $V$ while still providing 50% modulation for linear polarization in Stokes $Q$ and $U$. Because measuring circular polarization is our top priority, we maximize the modulation efficiency for Stokes $V$ and tolerate a reduced efficiency in Stokes $Q$ and $U$.

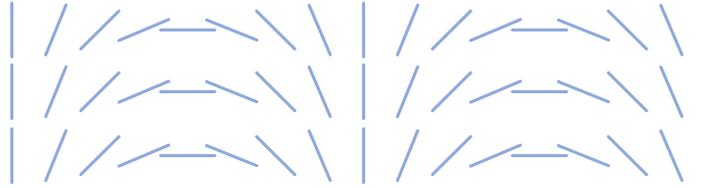

Figure 2. A spatially variable retarder where the fast-axis (indicated with blue lines) orientation angle changes linearly in the horizontal axis. This is the spatial equivalent of a rotating retarder.

A patterned retarder based on liquid-crystal technology where the fast axis rotates continuously along one dimension and is constant in the other dimension (see Figure 2) has distinct advantages:

- Coupled with a linear polarizer, the measured intensity signal $I'$ along the modulated dimension is the same as a polarimeter based on a rotating quarter-wave retarder.
- The polarization modulator has a wide range of acceptance angles because it is a true zero-order retarder.
- Retardation errors cannot create linear to circular cross-talk since the retardation only influences the amplitudes of the modulation and not the frequencies.
- The wavelength-dependence is proportional to the inverse of the wavelength since it is a true zero-order retarder. Multi-layer liquid crystals can substantially reduce the wavelength dependence of the retardation[7].
- The circular polarization Stokes $V$ is modulated at half the spatial frequency of the linear polarization (Stokes $Q$ and $U$). It is this difference in modulation frequencies that makes the approach extremely sensitive to circular polarization even in the presence of large amounts of linear polarization.
- Random errors in the fast-axis orientation only lead to a minor reduction in the efficiency with which the polarization can be measured (see Figure 3).
- Random errors in the fast-axis orientation can lead to cross-talk from linear to circular polarization but it can be neglected for typically manufacturing errors of 1 degree rms (see Figure 3).

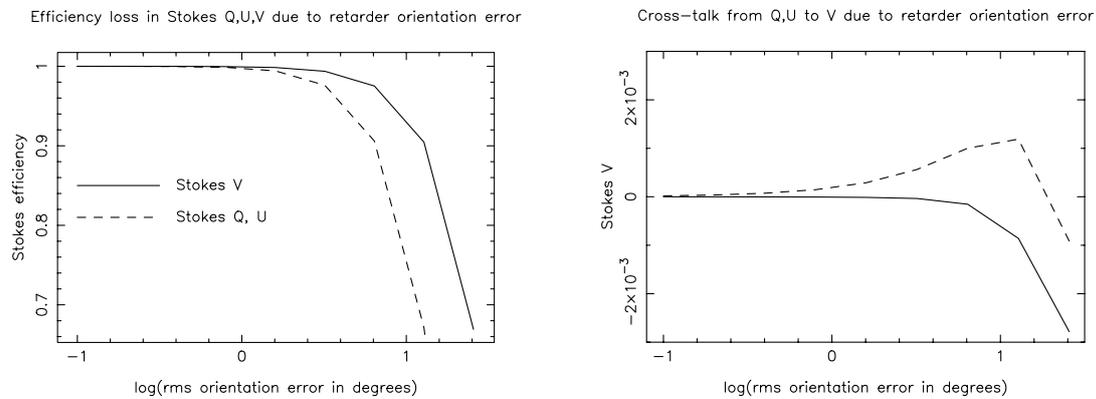

Figure 3. Influence of random fast-axis orientation errors on efficiency of polarization measurement (left) and cross-talk from linear (solid line is Q to V, dashed line is U to V) to circular polarization (right).

The polarization modulator approach also has some disadvantages that need to be taken into account when using it as the centerpiece of an instrument:

- Flatfield errors on the detector are a major limitation[3]. To reach $10^{-4}$ in circular polarization with 1024 pixels, random flatfield errors should not exceed $10^{-3}$ rms, which is difficult to achieve.
- The illumination along the modulation dimension needs to be uniform or independently measurable.

## 2.2 Spectroscopy

Since polarimetry is first, spectroscopy has to adapt to polarimetry being done by a patterned-retarder plate in the aperture, which must be reimaged onto the detector. Since 1) a spectrograph entrance slit also needs to be reimaged onto the detector, 2) we do not want any optical element in front of the polarization modulator, 3) the modulator needs to be illuminated uniformly, 4) and we do not (yet) require any angular resolution, we decided for a rather unconventional approach to spectroscopy (another reason to call it LSDpol) and make the entrance slit also the entrance aperture of LSDpol. In a classical spectrograph the entrance slit is in the image plane. Exchanging the focal and pupil planes as compared to normal spectrograph designs provides the highly uniform illumination needed for the required highly sensitive polarimetry.

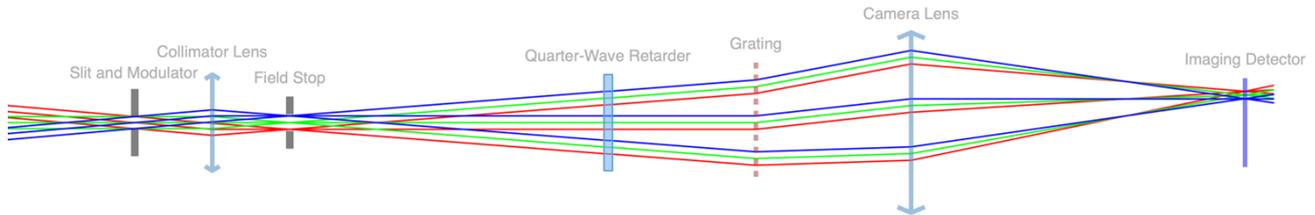

Figure 4. Paraxial layout of the LSDpol concept. Ray colors indicate three different fields. Only a single grating order is shown. The field angles are massively exaggerated to clearly show the three fields.

Our first optical design concept was based on the following considerations:

1. The spatial polarization modulator on its own cannot do polarimetry or spectroscopy. A polarization grating is an attractive element since it acts both as a dispersing element as well as a (circular) polarizing beam-splitter. By measuring two orthogonal polarization states simultaneously, we can add them up to measure the entrance slit illumination independent of the source's polarization.

2. Without a lens between the slit and the grating, different field points (different colors in image) would have different angles of incidence on the dispersing element, which would shift spectra on the detector depending on the field angle. We therefore need to exchange angles and location with a lens after the entrance slit.

3. By placing a (collimator) lens one focal length behind the entrance slit, the grating will be in a telecentric beam, which ensure that all field points will enter the grating with the same (cone of) angles.

4. By increasing the focal length of this collimator lens, the cone of angles can be reduced to increase the spectral resolution.

5. The size of the grating illuminated by a single field point will be less than the slit width if the grating is placed before the focus of the collimator lens, which will lead to insufficient spectral resolution due to the small number of illuminated grating rules.

6. If the grating is placed after the focus of the collimator lens, each field point can illuminate an arbitrarily large area of the grating

7. The focus of the telecentric beam provides the opportunity to place an internal field stop <u>after</u> the slit and thereby provide a well-defined, potentially adjustable field of view.

8. Since the spatial polarization modulator in the entrance slit also acts as a slight polarizing beam-splitter for circular polarization, the internal field stop will select slightly different fields of view for right- and left-handed circular polarization. This can be avoided with an external field stop or baffle in front of the modulator.

9. The telecentric beam is also the best place to place any flat optical element that is sensitive to the angle of incidence such as the quarter-wave retarder that makes the polarization grating into a polarizing beam-splitter for linear polarization.

10. A camera lens after the grating reimages the slit onto the detector. The reimaging is not perfect, but for the small field of view considered here, the corresponding reduction in spectral resolution can be tolerated.

11. If the camera lens is placed one focal length behind the grating, the telecentric beam will ensure that the linear dispersion will not depend on the detector focus.

The focal lengths of the two lenses determine the magnification. A reflective grating would provide a simple way to fold the beam and potentially make the instrument more compact. The slit width, field-of-view and spectral resolution all depend on each other, and we used a spreadsheet to optimize a paraxial design before analyzing a realistic design.

## 3. FIRST PROTOTYPE

### 3.1 Spatial Polarization Modulator

The spatial modulator was manufactured by ImagineOptix based on our specifications using a single layer of liquid-crystal material. Since we did not know what spatial modulation frequency would work best, our modulator has 10 different frequencies that increase by a factor of two from one to the next pattern. Figure 5 shows the orientation pattern and the manufactured modulator between crossed polarizers. Since the modulator is collocated with the entrance slit, the modulator can be moved perpendicular to the slit to select the desired modulation frequency.

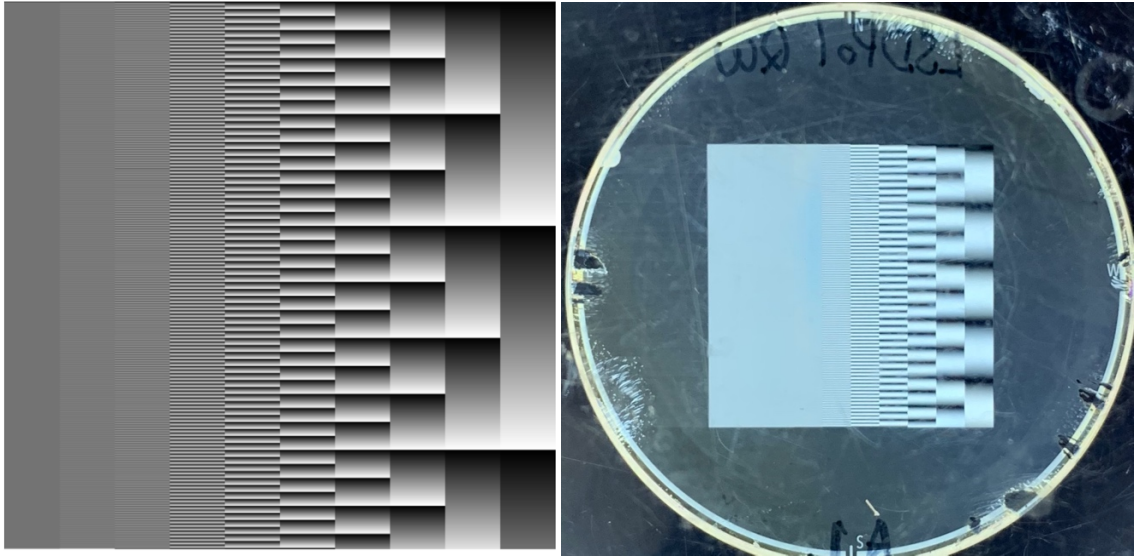

Figure 5. Orientation angle of fast axis spatial polarization modulator as designed (left) and manufactured modulator between crossed polarizers (right).

### 3.2 LSDpol Prototype

The prototype was assembled from off-the-shelf components and a few custom elements. The light path will pass the optical components in the following order:

1. Optional custom baffle(s)
2. Custom polarization modulator as described in the previous subsection
3. 0.1mm by 10mm custom entrance slit
4. Edmund Optics #32-724 60-mm focal-length lens
5. 4-12mm Thorlabs SM1D12CZ adjustable diaphragm as field stop
6. Edmund Optics #39-038 550-750 nm achromatic quarter-wave plate
7. Edmund Optics #12-678 286 grooves/mm polarization grating
8. Edmund Optics #49-662 30-mm focal-length lens
9. Basler camera with SONY CMOS IMX249 detector

Initially the components were mounted in individual optics mounts for easy adjustment and exchange of lenses. In a second phase, the components were mounted inside a cage system so that the LSDpol prototype could also be used in the field.

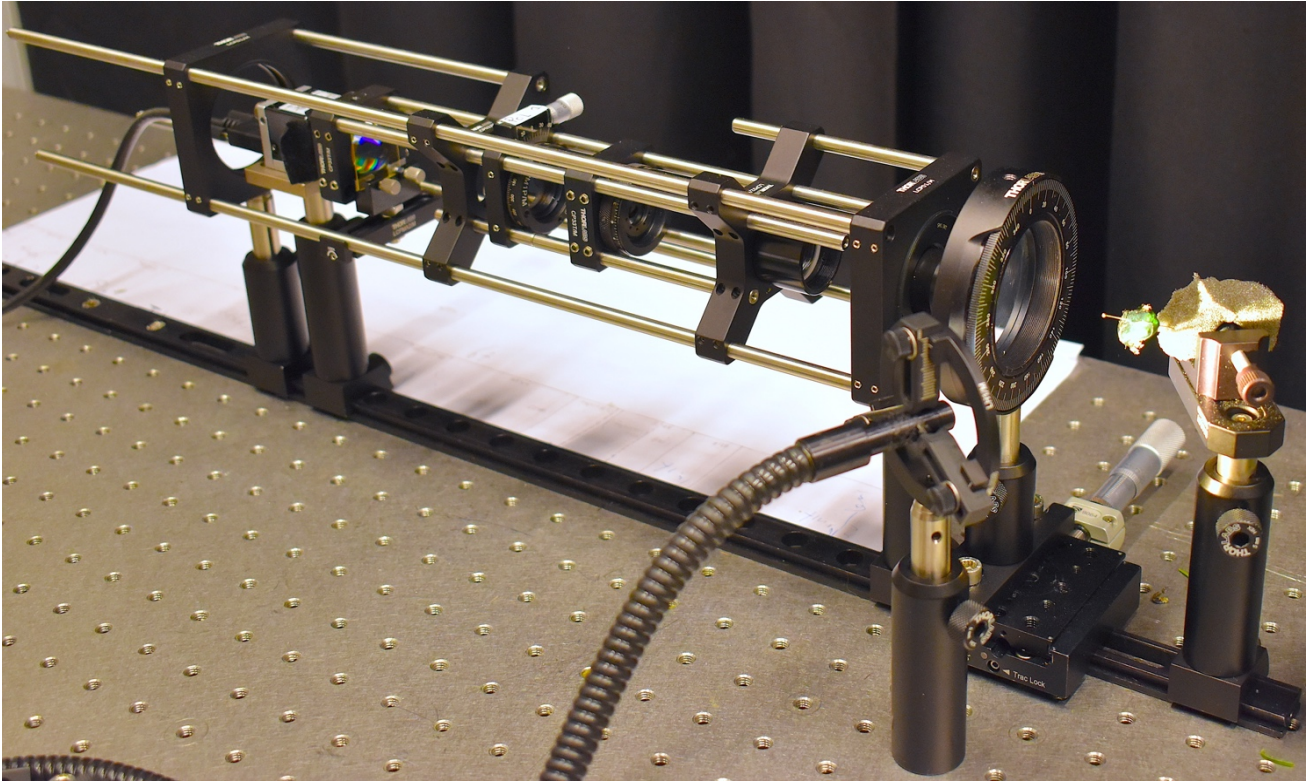

Figure 6. LSDpol prototype measuring circularly polarized reflection spectra from a green beetle.

## 4. PRELIMINARY DATA ANALYSIS

### 4.1 Calibration

Before data can be analyzed, LSDpol needs to be calibrated. The conversion from pixels to wavelengths can easily be determined from an image of a spectral-line lamp. A single image from a 100% linearly polarized source such as shown in Figure 7 is sufficient to determine all parameters that are relevant for polarimetry:

1. Determine the background signal (bias, dark current, and straylight) from the darkest parts of the images, which occur between the zeroth and the first order spectra.
2. Determine the mutual alignment of the left and the right spectra knowing that the modulation signals should have opposite signs.

The following steps are carried out for each wavelength separately.

3. Determine the modulation amplitudes in the left and the right spectra from the rms intensity along the slit and knowing that the modulation is sinusoidal.
4. Determine the polarization modulator retardation $\delta$ knowing that the measured modulation amplitude for Stokes $Q$ and $U$ is $\frac{1}{4}(1 - \cos\delta)$.
5. Since Stokes $Q$ has a mean value of $\frac{1}{2} + \frac{Q}{4}(1 + \cos\delta)$ and Stokes $U$ always has a vanishing mean, we can determine the amount of $Q$ and $U$ in the incoming light as long as we know that the input is 100% linearly polarized. We do not need to know the angle of the linear polarization that we use to calibrate LSDpol.
6. The spatial polarization modulation frequency and phase is estimated from a Fourier transform of the signal along the slit. A better phase estimate is obtained from a cross-correlation with a sinusoidal function with the frequency determined from the Fourier transform.

7. Due to optical distortions, the modulation frequency as seen by the camera is not constant along the slit and varies with wavelength. We found an easy approach to correct for that by locally determining the phase shift between a perfectly constant and the actually varying modulation. If the phase shift is smaller than about 1rad, the following equation, based on the observed modulation $I_{obs}$ and a Taylor expansion of the simulated modulation $I_{sim}$ and a weighted mean of the derivative of the simulated modulation to avoid divisions by zero can be used to determine the local shift:

$$\Delta\theta = (I_{obs} - I_{sim}) \cdot \frac{\partial I_{sim}}{\partial \theta} \cdot \frac{1}{\langle \left(\frac{\partial I_{sim}}{\partial \theta}\right)^2 \rangle}$$

## 4.2 First Test Measurements

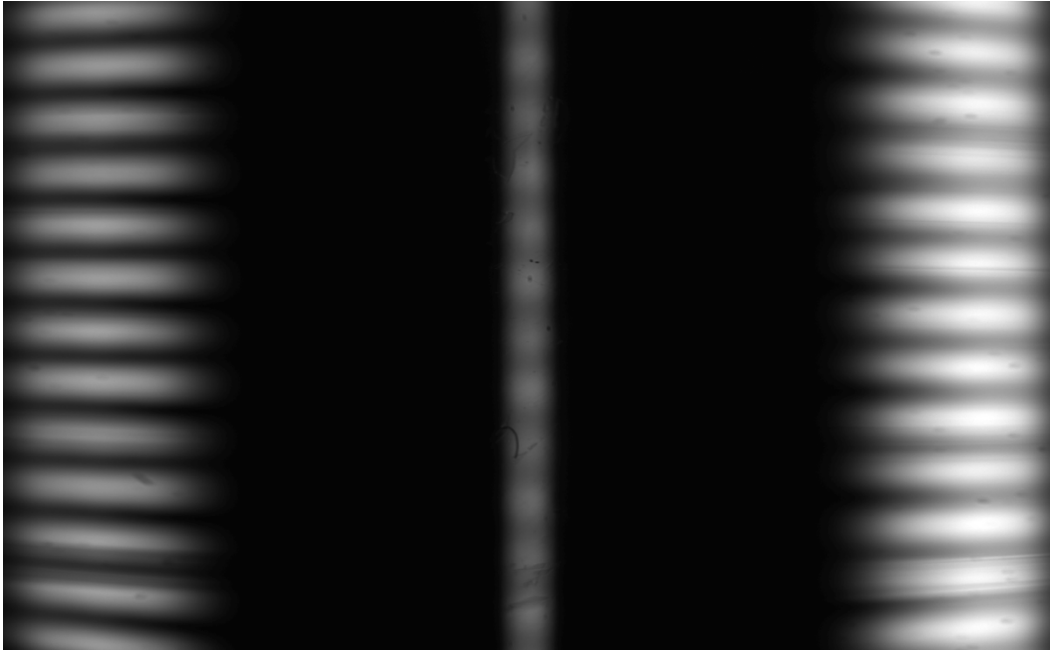

Figure 7. Raw linear-polarizer calibration image from the LSDpol prototype. The zeroth grating order (m-0) is in the center, opposite linear polarization states are on the left (m=-1) and the right (m=1). Note the barrel distortion towards the edges of the image and the spurious modulation in the zeroth-order slit image.

We extensively tested the prototype and also added a housing that makes it suitable for field testing. Figure 7 shows a raw image from a calibration with a linear polarizer. Issues such as distortion and spurious signals will be discussed in the following section. Figure 8 shows a measurement of unpolarized light reflected by a beetle; note that the modulation occurs at half the frequency of the linear calibration polarizer, indicating a strong circular polarization signal.

## 4.3 Data Reduction

We normalize the left spectrum with the sum of the left and right spectrum and only work with this single spectrum, $I_{obs}$. The normalized right spectrum contains no additional information since it is identical to 1 minus the left normalized spectrum. Flatfield defects are identified and masked before the polarization is determined via a fixed-phase demodulation:

$$\frac{Q'}{I'} = \frac{1}{4}(1 - \cos\delta')\cos 4\theta', \qquad \frac{U'}{I'} = \frac{1}{4}(1 - \cos\delta')\sin 4\theta', \qquad \frac{V'}{I'} = \frac{1}{2}\sin\delta'\sin 2\theta',$$

where $Q'/I'$, $U'/I'$ and $V'/I'$ are the estimated normalized Stokes parameters of the source; $\theta'$ and $\delta'$ are the estimated modulator fast axis rotation and retardation, respectively, as determined with the linear-polarizer calibration.

## 4.4 Example Measurements

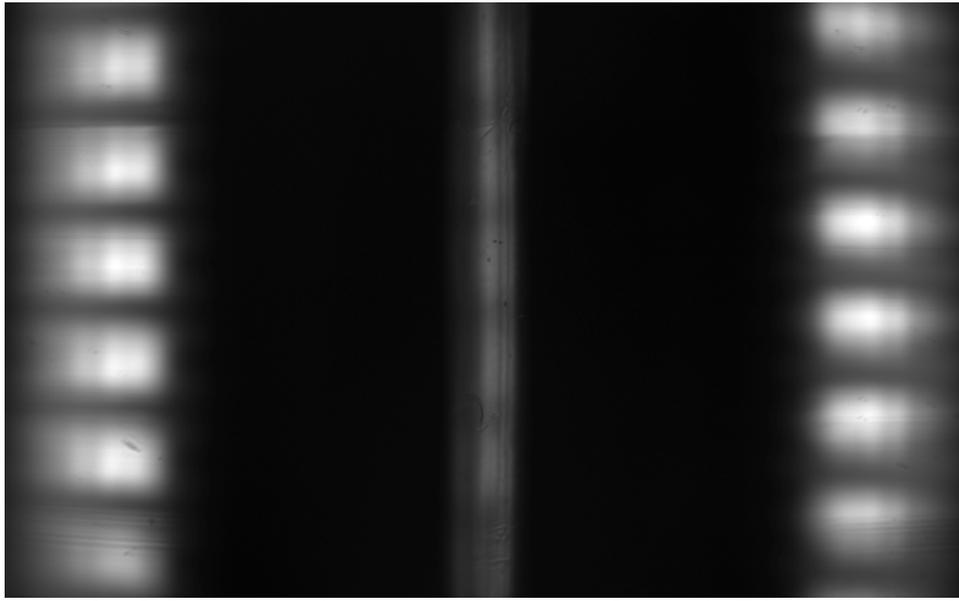

Figure 8. Raw image from LSDpol prototype looking at a green beetle in reflected light. This particular type of beetle is known to create a large amount of circular polarization when illuminated with unpolarized light.

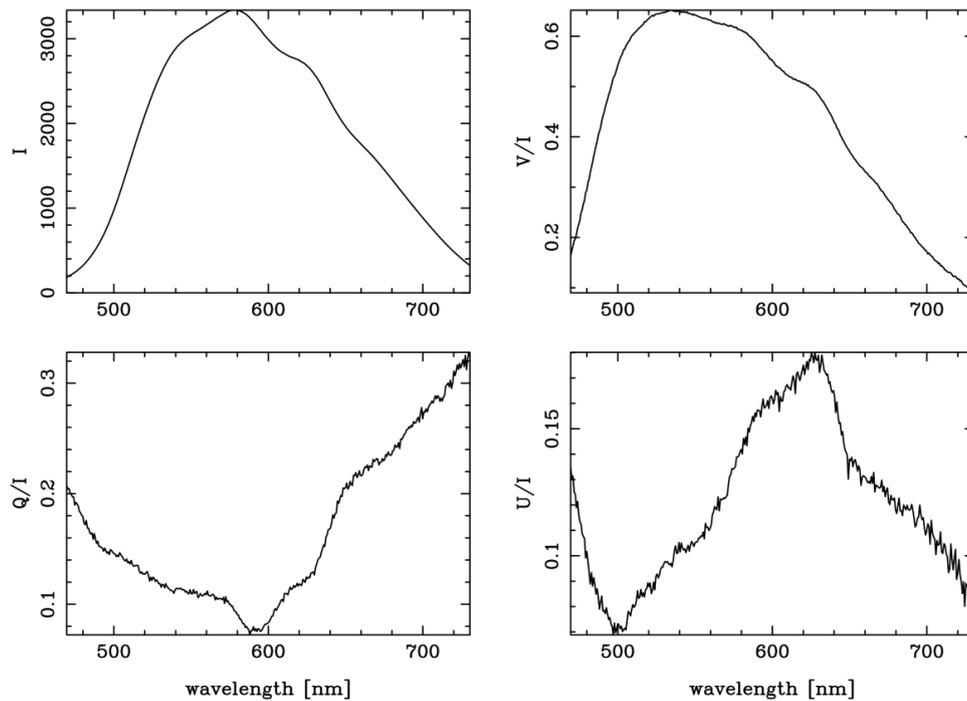

Figure 9. Linear and circular polarization of light reflected by a green beetle. Note the large fraction of circular polarization exhibited in Stokes V/I (top right). The linear polarization signals contain a substantial amount of cross-talk from circular to linear polarization and can therefore not be trusted.

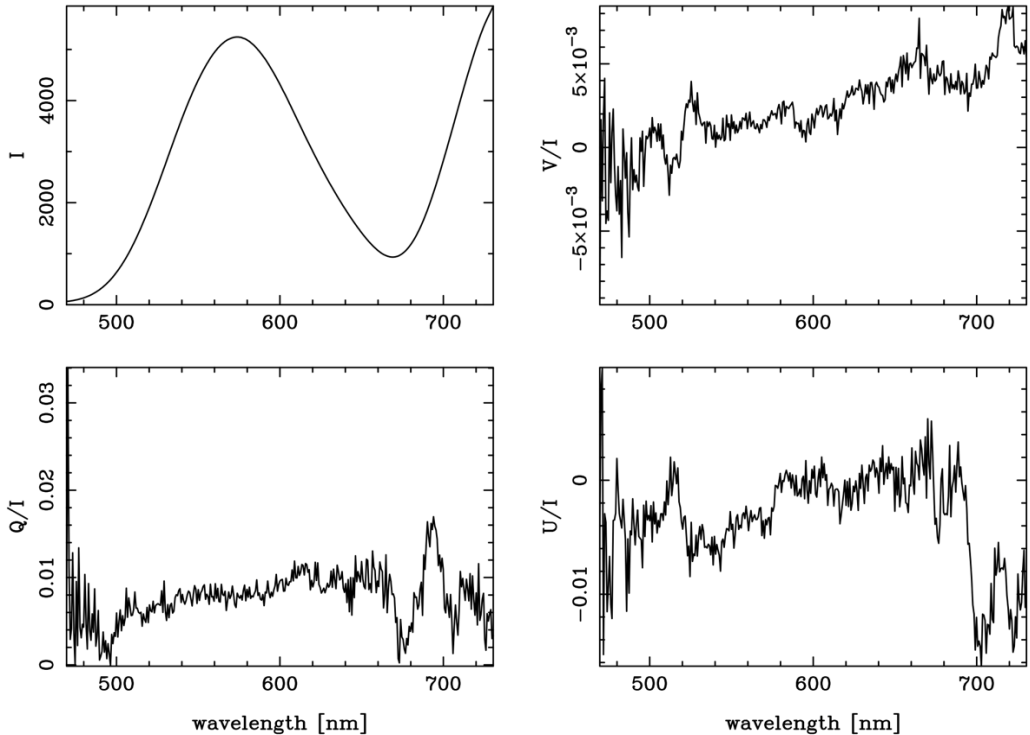

Figure 10. Linear and circular polarization of light reflected by a green leaf. No clear signal is visible, although there are indications for substantial variations around 700 nm.

The beetle data were then reduced as described above. Figure 9 shows the uncalibrated intensity (not corrected for the wavelength-dependence of the transmission and the detector efficiency) along with the normalized linear and circular polarization signals. There is a clear circular polarization signal that goes above 60% in the green part of the spectrum. The random noise is much less than 1% rms per wavelength. As such, it should be possible to detect small circular polarization signals at the $10^{-4}$ level by averaging over wavelength bins. However, attempts to reliably measure very small circular polarization signals <0.1% were not yet successful due to systematic errors that will be discussed in the following section. An example of light reflected from a green leaf is shown in Figure 10.

## 5. LESSONS LEARNED

As is so often the case, practice deviates significantly from theory. There is only one major issue that we encountered, and it can be fixed in a future LSDpol implementation:

- Our approach to polarizing beam-splitting with a quarter-wave plate and a polarization grating is elegant, but introduces linear to circular cross-talk. While the fast-axis orientation of the quarter-wave plate only changes the phase of the modulation, the retardation deviation from a perfect quarter-wave plate leads to a wavelength-dependent cross-talk from linear to circular polarization. Although we use an achromatic quarter-wave retarder, even minor deviations lead to substantial crosstalk. The retardation of the quarter-wave plate in front of the polarization grating can, in principle, be determined from the amplitude of the apparent Stokes *V* modulation at half the frequency of the linear polarization introduced by the calibration polarizer. However, this approach has not (yet) been successful at sufficiently reducing the crosstalk during the data reduction. We will therefore explore other approaches to dispersing and polarizing beam-splitting, which would also take care of the limited wavelength range of the polarization grating (450nm-650nm) and thereby provide much better performance in the 650nm-750nm wavelength range where green plants shows the largest circular polarization signals.

Other, less fundamental issues that also should be addressed in the next version of LSDpol include:

- The spatial polarization modulator has defects (due to mishandling, not manufacturing) that are reimaged onto the detector and influence the calibration and the measurements. They are difficult to remove because they influence the polarization, not just the intensity. Future implementations will use the second spatial modulator that we have.

- The flatfield is rather bad because of dirt on the camera. We used elaborate outlier detection algorithms to remove the influence of dirt. Future implementations need to make sure that the detector remains clean. In addition, the linear polarization calibration should be performed at several angles so that the modulation and the flatfield can be separated. This should allow us to obtain a good flatfield.

- Distortion is a major issue, in particular because it is not completely symmetric with respect to the spectra on the left and the right. Hence, adding and subtracting the two polarization states to remove the slit illumination and transmission variations is only partially successful. Future implementations need to improve on the camera lens to reduce the distortion.

- The instrument is rather lengthy at about 50 cm. We would like to reduce that to less than 20 cm so that it fits into a 2U or 3U CubeSat.

- The spatial polarization modulator also acts as a weak grating, which might also explain the spurious modulation seen in the slit image in the zeroth grating order. Furthermore, it might also explain a spurious modulation at the 1% level with a frequency of $2\theta$ like Stoke $V$ but phase-shifted by 90 degrees. A full diffractive model including polarization needs to be established to assess the impact of this grating.

## ACKNOWLEDGEMENTS

We thank Sujeeporn Tuntipong for assembling the very first version of LSDpol. The research of Frans Snik and David S. Doelman leading to these results has received funding from the European Research Council under ERC Starting Grant agreement 678194 (FALCONER).